\documentclass[useAMS,usenatbib, usegraphicx, epsfig]{mn2e}

\usepackage{amssymb}
\usepackage{subfigure}
\usepackage{amsmath}
\usepackage{multirow}

\title[Tomographic Magnification]{Tomographic Magnification of Lyman Break Galaxies in The Deep Lens Survey}

\author[C.~B.~Morrison et al.]{C.~B.~Morrison$^{1}$\thanks{E-mail:cbmorrison@ucdavis.edu},
R.~Scranton$^{1}$,
B.~M\'enard$^{2,3}$\thanks{Alfred P. Sloan Fellow},
S.~J.~Schmidt$^{1}$, 
J.~A.~Tyson$^{1}$,
\newauthor
R.~Ryan$^{4}$, 
A.~Choi$^{5}$, 
D.~M.~Wittman$^{1}$ \\
$^{1}$Department of Physics, University of California, One Shields Avenue, Davis, CA 95616, USA \\
$^{2}$Department of Physics and Astronomy, Johns Hopkins University, 3400 N. Charles Street, Baltimore, MD \\
$^{3}$Institute for the Physics and Mathematics of the University of Tokyo, Kashiwa 277-8583, Japan \\
$^{4}$Space Telescope Science Institute, 3700 San Martin Drive, Baltimore, MD 21218, USA\\
$^{5}$Scottish Universities Physics Alliance, Institute for Astronomy, University of Edinburgh, Royal Observatory, Blackford Hill, Edinburgh EH9 3HJ}
\begin{document}

\maketitle

\begin{abstract}
Using about 450,000 galaxies in the Deep Lens Survey, we present a detection of the gravitational magnification of $z>4$ Lyman Break Galaxies by massive foreground galaxies with $0.4<z<1.0$, grouped by redshift. The magnification signal is detected at S/N greater than 20, and rigorous checks confirm that it is not contaminated by any galaxy sample overlap in redshift. The inferred galaxy mass profiles are consistent with earlier lensing analyses at lower redshift. We then explore the tomographic lens magnification signal by splitting our foreground galaxy sample into 7 redshift bins. Combining galaxy-magnification cross-correlations and galaxy angular auto-correlations, we develop a bias-independent estimator of the tomographic signal. As a diagnostic of magnification tomography, the measurement of this estimator rejects a flat dark matter dominated universe at $>7.5\sigma$ with a fixed $\sigma_8$ and is found to be consistent with the expected redshift-dependence of the WMAP7 $\Lambda$CDM cosmology. 
\end{abstract}

\begin{keywords}
gravitational lensing: weak---large-scale structure of the Universe---cosmology: observations---galaxies: high-redshift---galaxies: haloes
\end{keywords}

\section{Introduction}

Lensing tomography is a powerful probe of cosmology. By comparing the amplitude of lensing effects across multiple redshift baselines, we can observe both the geometric expansion of the universe and the growth of structure as a function of time.  These are two key diagnostics of dark energy, making tomography an essential part of the toolkit identified by the Dark Energy Task Force \citep{Albrecht06} for use in upcoming sky surveys.

Gravitational lensing has two observational signatures: shear and magnification. While gravitational shear has been the main focus of observational efforts in the past decade, magnification measurements are now possible with the availability of deeper surveys together with stable photometry. Weak magnification by galaxies robustly detected by \citet{Scranton05} based on the cross-correlation between low redshift galaxies with distant quasars from the Sloan Digital Sky Survey (SDSS). Subsequent work by \citet{Menard10} showed that shear and magnification measurements give consistent mass profiles around galaxies. Since then, the detection of gravitational magnification has been reported in various contexts: using Lyman Break Galaxies (LBGs) \citep{Hildebrandt09b}, sub-millimeter galaxies \citep{Wang11}, galaxy groups \citep{Ford12}, and clusters \citep{Hildebrandt11}. Several estimators have been used, based on source number density \citep{Scranton05}, brightness change \citep{Menard10}, size change \citep{Huff11,Schmidt12} and quasar variability \citep{Bauer11}.

So far attempts to detect a tomographic lensing signal have only been made through shear measurements \citep{Schrabback10}.
Such analyses require accurate shape measurements as a function of redshift. In contrast, magnification measurements do not rely on shape measurements and can even be done for unresolved sources. In this paper, we use LBGs at redshift $z\sim4$ as background sources and cross-correlate with foreground lenses. We then extend the magnification measurement, breaking the foreground lenses into bins of photometric redshift.  By selecting multiple photometric redshift samples, we  probe the growth of structure and distance for several different ranges of cosmic time, albeit at the expense of having galaxy samples with redshift-dependent biases.  To account for this bias, we normalize each of the lensing measurements by the observed foreground sample autocorrelation, resulting in a signal that is nearly bias independent \citep{Jain03, VanWaerbeke10}.

The paper is structured as follows: in \S\ref{sec:theory} we derive the magnification formalism and the expected tomographic signal; \S\ref{sec:data} covers the details of the Deep Lens Survey (DLS), the data selection criteria, and the analysis techniques used; \S\ref{sec:results} describes the results; and in \S\ref{sec:conclusions} we summarize and discuss future directions. We discuss detailed survey parameters, possible systematics, and theoretical modeling of the expected signal in the Appendix.

All cosmological calculations assume, unless otherwise stated, a flat universe with $\Omega_{\Lambda}=0.73$, $\Omega_m=0.27$, and $\sigma_8=0.8$. Magnitudes and color cuts are done using $AB$ magnitudes.

\section{Theory}\label{sec:theory}

Gravitational magnification can change the apparent density of background sources \citep{Narayan89} and as a result 
induce apparent cross-correlations between foreground and background galaxy populations. In this section we present the formalism describing magnification-induced spatial correlations following \citet{BartelmannSchneider01}.

For a flux-limited sample, the integrated number of galaxies above a given flux threshold $f$ can be approximated as
\begin{equation}
N_0(>f)\sim A f^{-\alpha}
\end{equation}
where $N_0$ is the number of galaxies in area $A$ and $\alpha$ is the power-law slope of the number counts. Magnification increases the depth observed by reducing the effective flux limit to $f\rightarrow f/\mu$ and simultaneously decreasing the area of the survey by $A \rightarrow A/\mu$, where $\mu$ is the magnification. Combining these effects, the relation for the galaxy number counts becomes
\begin{equation}
N(>f)\sim\frac{1}{\mu}A \left(\frac{f}{\mu}\right)^{-\alpha}=\mu^{\alpha-1}N_0(>f)\label{eq:mag_flux}
\end{equation}
From this functional form, we see that for values of $\alpha>1$ the number of sources observed is increased while the opposite is true for $\alpha<1$. At the critical value of $\alpha=1$, no effect from magnification is observed.

We introduce the cross-correlation between lenses and sources as follows: we can define the over-density of sources $N_s$ as a function of position on the sky $\phi$ to be
\begin{equation}
\delta_{s}(\phi) = \frac{N_s(\phi)-\langle N_s \rangle}{\langle N_s \rangle}.
\end{equation}
where $\langle N_s \rangle$ is the average density. Under the effects of magnification given in Equation~\ref{eq:mag_flux}, the number of sources is
\begin{equation}
N_s(\phi) = \mu(\phi)^{\alpha-1} \langle N_s \rangle.
\end{equation}
We consider small departures of $\mu$ from unity and substitute $\mu=1+\delta_\mu$ where $|{\delta_\mu}| \ll 1$. Taylor expanding, we can write $\mu^{\alpha-1}\approx 1+(\alpha-1)\delta_\mu$, thus the over-density under magnification becomes
\begin{equation}
\delta_{s}(\phi) = (\alpha-1)\delta_{\mu}.
\end{equation}
Hence, the angular cross-correlation between lenses and sources induced by magnification can be written as
\begin{eqnarray}
\label{eq:mag_correlation}
w_{ls}(\theta) &=& (\alpha-1) \langle\delta_\mu(\phi)\delta_g(\phi+\theta) \rangle \nonumber \\
 &=& (\alpha-1) w_{\mu \delta_g}(\theta)
\end{eqnarray}
where $\theta$ is the angle from the lensing galaxy center and the foreground galaxy density is $\delta_g$. We use Limber's approximation to convert the intrinsically 3D distribution of galaxies to projected angle. We do this by expanding the above equation in the same framework as in \citet{BartelmannSchneider01}. We then have

\begin{eqnarray}
w_{\mu \delta_g}(\theta) &=& \int^{\chi_H}_0d\chi \, \eta_l(\chi) K(\chi) \nonumber \\ &~& \times \int^{\infty}_0 \frac{kdk}{2\pi} b(k, \chi) r(k, z) P_{\rm DM}(k,\chi)J_0(\chi k\theta) \label{eq:mag_corr}
\end{eqnarray}
where $\chi$ is the co-moving distance, $P_{\rm DM}$ is the dark matter power spectrum, b is the galaxy bias as a function of scale and distance, r is correlation coefficient also as a function so scale and distance, $\eta_l$ is the co-moving distance distribution of the foreground lensing galaxies, $J_0$ is the zeroth order Bessel function, and
\begin{equation}
K(\chi)=\frac{3 H^2_0\Omega_m}{c^2} \frac{\chi}{a} \int^{\chi_H}_\chi d\chi^\prime \eta_s(\chi^\prime)\frac{\chi^\prime-\chi}{\chi^\prime} \label{eq:mag_kernel}
\end{equation}
is the lensing kernel-weighted distribution of background sources $\eta_s(\chi)$.

For the remainder of the paper, we make the assumption that the galaxy bias is linear, that is $b(k, \chi) \rightarrow b$, a constant, and $r(k, \chi) \rightarrow 1$. There are limitations to these assumptions. If the bias is stochastic, $r<1$, and the bias is a strong function of $k$ then interpreting these measurements cosmologically will be difficult. There is observational evidence that this is the case \citep{Hoekstra02, Simon07}, however a recent measurement in COSMOS \citep{Jullo12} found that for galaxies selected in photometric redshift, at redshifts similar to this analysis, $b$ was linear to small scales and $r=1$ to within the measurement error. We attempt to mitigate these effects in the next section and show how well we can for realistic HODs, similar to our own.

\subsection{Tomographic Signal}\label{sec:tomographic_theory}

The galaxy bias $b$ complicates the interpretation of magnification results with regard to cosmology and large scale structure, since the bias is dependent on galaxy type, brightness, and mass -- any or all of which may vary in the lensing sample as a function of redshift.  If we want to isolate the cosmological evolution of the lensing signal, we have two choices.  Either we can model the galaxy bias for each redshift bin and marginalize over the parameters for that model or we can try an entirely empirical approach where we attempt to cancel out the bias using the autocorrelation of the foreground sample by constructing a bias independent (or nearly bias independent) estimator.  For the purposes of this paper, we choose the latter.

Using the same assumption that the density of galaxies follows the dark matter over-density as $b \delta$, the autocorrelation for the lensing galaxies is given by
\begin{equation}
w_{ll}(\theta) = b^2 \langle \delta(\phi) \delta(\phi+\theta) \rangle.
\end{equation}
Using Limber's approximation, this becomes
\begin{eqnarray}
w_{ll}(\theta) &=& b^2 \int^{\chi_H}_0 d\chi \, \eta_l(\chi)^2 \nonumber \\ &~& \times \int^{\infty}_0 \frac{kdk}{2\pi} P_{\rm DM}(k,\chi)J_0(\chi k \theta)
\end{eqnarray}
where the definitions follow from the previous section. In this linear approximation of the galaxy bias, we can remove the bias by taking an appropriate ratio of the magnification and autocorrelation signals:
\begin{equation}
\label{eq:bias_free}
\mathcal{R} = \frac{w_{ls}^{2}}{w_{ll}}.
\end{equation}
This ratio is independent of galaxy bias, as long as the linear approximation of the bias holds.

Observationally, we do not directly measure $w_{ll}$, instead we measure the true autocorrelation with an additional constant value. Pair conservation introduces a measurement bias $C$:
\begin{equation}
\label{eq:int_const}
C(\theta) = [1+w_{ll}(\theta)]N^{-2}\sum_{i} w_{ll}(\theta_i)
\end{equation}
for $N$ galaxies in the survey area with the sum taken over all pairs. In practice, $C(\theta)$ is nearly constant as function of scale, resulting in an overall suppression of $w_{ll}$ \citep{Peebles80, Scranton02}.  We use Equation~\ref{eq:int_const} to correct the amplitudes of the measured auto-correlation in the correlation ratio, giving
us a final form of
\begin{equation}
\label{eq:bias_ratio}
\hat{\mathcal{R}}= \frac{(w_{ls})^2}{w_{ll}-C}.
\end{equation}

We model Equation~\ref{eq:bias_ratio} in the quasi-linear and non-linear regime, and also account for the effects of the integral constraint on the autocorrelation measurements.  For this purpose, we use the code {\it NICAEA}\footnote{Kilbinger, Martin: http://www2.iap.fr/users/kilbinge/nicaea/} to generate the dark matter power spectrum, $P_{DM}$, and subsequent correlations.

\begin{figure}
\includegraphics[width=0.495\textwidth]{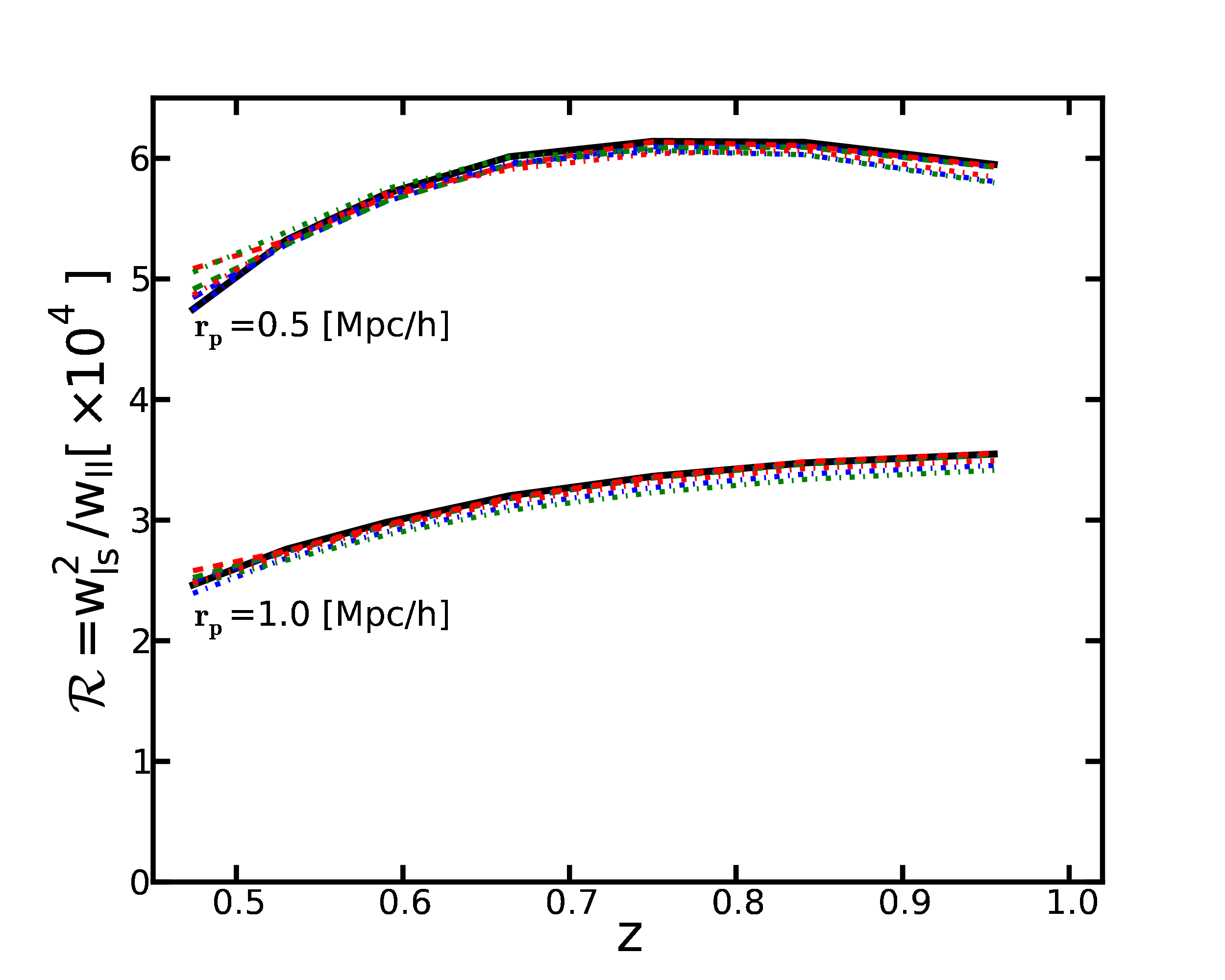}
\caption{\label{fig:hod_compare}
Predicted curves for the ratio $\mathcal{R}(z)$ on 0.5 and 1.0 Mpc/$h$ scales, for a range of models.  The solid black line is the dark matter-only ratio, dashed curves use HODs derived from DLS galaxy-galaxy lensing at $z \sim 0.6$ (private communication A. Choi), and dot-dashed curves are fits to galaxy clustering at $z \sim 1.1$ from \citet{Wake11}.  For both HOD classes, the red, green and blue curves run over a range of galaxy bias.  See Table \ref{tab:hod_models} for more details.}
\end{figure}

To test the bias independence of the quantity $\hat{\mathcal{R}}$ at small scales we utilize the halo model as defined in \citet{Seljak00}. To estimate the galaxy power spectrum, $P_{gg}$, and galaxy-matter cross power spectrum, $P_{gm}$ we use two models for populating galaxies in halos. The first is the halo occupation distribution (HOD) from \citet{Zheng07} with model parameters from \citet{Wake11}. For the second we use the HOD from \citet{Mandelbaum05} with parameters as measured in the DLS from galaxy-galaxy lensing (private communication A. Choi). We chose these parameters as they bracket this analysis both in redshift and halo masses. Figure~\ref{fig:hod_compare} shows the resultant ratio $\mathcal{R}$ at two physical distances, 1 Mpc/$h$ and 0.5 Mpc/$h$, for the two HODs compared to the dark matter. It should be noted that the redshift binnings used in this plot are not the same as those from the data, however, the resultant bias independence of the ratios will be similar. The ratio using HODs agree with the dark matter only ratio, at worst deviating by 20\%. The $\mathcal{R}$ and $\hat{\mathcal{R}}$ ratios are then bias-independent at the scales used in this analysis. For full details on the HODs and halo modeling see Appendix \ref{sec:hod_theory}.

\section{Data}\label{sec:data}

Reliably measuring weak lensing magnification demands accurate, consistent photometry over the whole survey area utilized. In this section we lay out the calibration steps we use to achieve this, and discuss the selection of both lenses and sources.

\begin{figure*}
\includegraphics[width=0.495\textwidth]{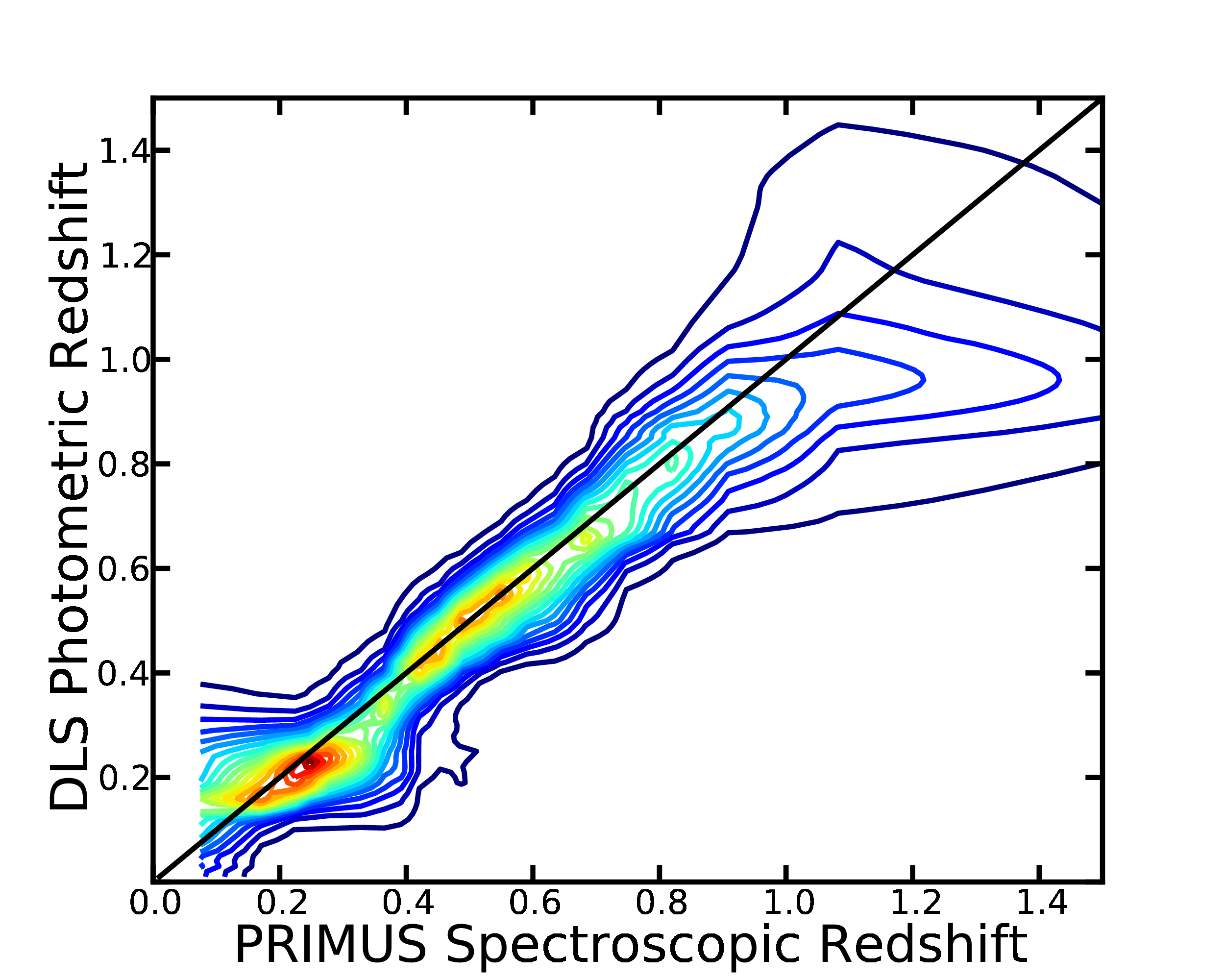}
\hfill
\includegraphics[width=0.495\textwidth]{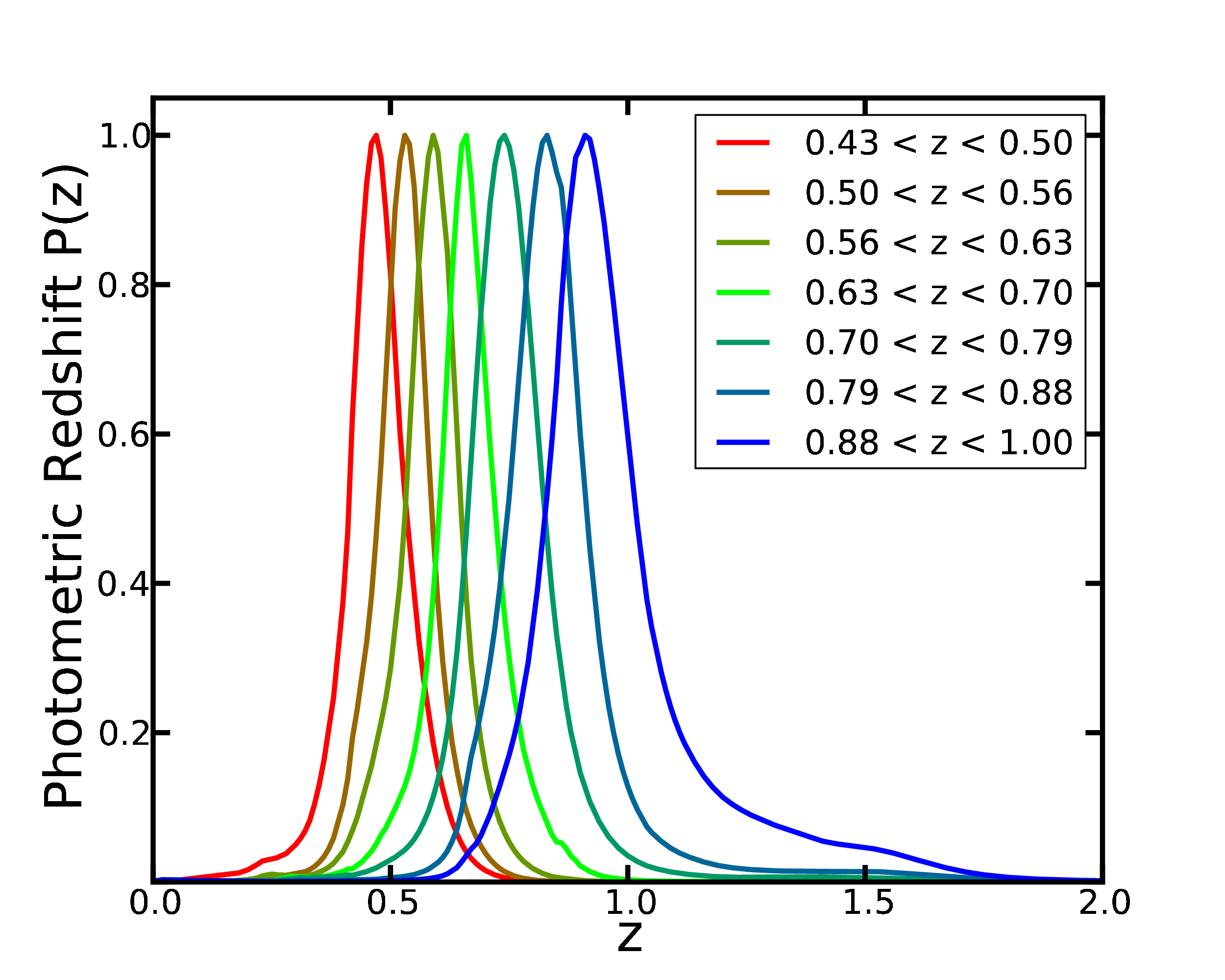}
\caption{\label{fig:photoz}
Plots of the DLS photometric redshift performance. The figure on the left shows the summed redshift posterior probability $p(z)$ vs spectroscopic redshift from  $\sim 9000$ PRIMUS survey galaxies binned by spectroscopic redshift. The contour values are linearly spaced and range from $200/dz^2$ to $3800/dz^2$ where dz is 0.01 in redshift . Integrating over the full range of spectroscopic and photometric redshifts gives the total number of galaxies in PRIMUS that overlap the DLS. The right panel shows the redshift distribution for the 7 photometric redshift-selected bins based on the aggregate $p(z)$ for the galaxies in each bin, weighted by the likelihood that a given galaxy is inside the bin. The curves are normalized to a maximum of one.
}
\end{figure*}

\subsection{The Deep Lens Survey}

The Deep Lens Survey (DLS) \citep{Wittman02} is a four band, $BVRz$, CCD survey of 20 square degrees observed on both the KPNO Mayall and CTIO Blanco telescopes, optimized for the observation of gravitational lensing. The survey is composed of five $4~ {\rm deg}^2$ fields (two northern and three southern), widely spaced in RA/Dec at high galactic latitude. The MOSAIC imagers \citep{Pogge98, Muller98} served as the observing instruments during the course of the survey, with each field consisting of a $3 \times 3$ grid of MOSAIC footprints for a total of 45 subfields. Nights with the best seeing (PSF FWHM $< 0.9\arcsec$) were reserved for $R$ band, leading to a total exposure time of 18,000 seconds. The remaining bands all have a total exposure time of at least 12,000 seconds. Because $V$ observations were often taken on good seeing nights after finishing $R$ observations, the seeing in $V$ band is also $\sim 1\arcsec$. The survey is 50 \% complete in terms of object recovery to $26_{AB}$ in $R$ band, $25.5_{AB}$ in $BV$ and $24.5_{AB}$ in $z$.

The DLS image processing and photometry are described in detail in an upcoming data release paper (Wittman et al. in prep.) as will be the photometric redshifts used in this analysis (Thorman \& Schmidt in prep.).  The details of these pipelines relevant to this analysis can be found in the Appendix.

\subsection{Foreground Lens Selection}\label{sec:lens_selection}

To measure magnification and tomography the lens sample must be free of spatially varying survey systematics (see Appendix \S\ref{sec:systematics} for more details) as well as have good photometric redshifts. We accomplish this by selecting galaxies with magnitudes in the range $20 \le m_R \le 24$, which is well below and well above the saturation and detection limits respectively. This cut also assures quality photometric redshifts due to the high signal to noise of the galaxies in this range.

We used the redshift posterior probability, $p(z)$, calculated by the Bayesian redshift estimator code, {\it BPZ} for each galaxy (see \S\ref{sec:photoz} for details) to define redshift bins. The posterior probability distribution for each galaxy is summed, yielding an estimate of the redshift distribution for the foreground lenses (albeit one that was convolved with the redshift-dependent scatter of the photometric redshifts).  We then split this distribution into 15 bins of equal likelihood over the range $0 < z < 5$.  Since the summed likelihood is a convolution of the true redshift distribution and the photometric redshift scatter, this binning produces broader redshift bins at higher redshift, as one would expect. The choice of 15 bins guarantees that no bin is narrower than $\delta z \sim 0.06$, the expected scatter in the DLS photometric redshifts.

From the initial binning, we select 7 bins that span the peak of the lensing kernel for a redshift $z=4$ (the redshift of the Lyman Break Galaxies, see \S\ref{sec:lbg_selection}) source.  The filter set of the DLS, $BVRz'$, allows for accurate redshifts within the range $z=0.4-1.0$, as such we only consider galaxies within this redshift range in the analysis.

To test how well the photometric redshifts are performing we use the spectroscopic survey PRIMUS (\citet{Coil11}, Cool et al. in prep.) as a cross check. The left panel of Figure~\ref{fig:photoz} shows the summed $p(z)$ versus spectroscopic redshifts from the PRIMUS survey that overlap the DLS. From PRIMUS we have $\sim9000$ spectroscopic redshifts which are 100\% complete to an $m_R$ band magnitude of $22.8_{AB}$ and 30\% complete to $m_R$ of $23.3_{AB}$. Contours of $p(z)$ in the redshift range $z=0.4-1.0$ track the one-to-one line in a mostly unbiased manner, with their expectation value within $z=0.02$ of the mean spectroscopic redshift value. There are also no significant spurious peaks in range of redshifts plotted, however, there are degeneracies between redshifts of $z \sim 2$ and $z<0.3$ caused by a lack of $U$ band data. We find that, for the range of redshifts used in this analysis, the full $p(z)$ is robust against catastrophic outliers and we find similar benefits to using the full $p(z)$ to those in \citet{Wittman09}. For the galaxies with spectroscopic redshifts from PRIMUS, the catastrophic outlier rate for point estimate redshifts defined as the peak of the redshift posterior is 7.6\%, and drops to 3.5\% in the $z=0.4-1.0$ range used in this analysis.

For each galaxy, we calculated the probability $P_i$ that it is within a given redshift bin $i$'s bounds $z_{min, i} < z < z_{max, i}$
\begin{equation}
P_i = \int_{z_{min, i}}^{z_{max,i}}p(z) ~ dz .
\end{equation}
To select that a galaxy be in a given bin, we require that $P_i > 0.16$, which is the single tailed $1\sigma$ probability of being in bin $i$.  To avoid double-counting, we then weight each galaxy by their $P_i$ value for a given bin.  This selection is similar to a photometric redshift ODDS cut, where a requirement is made that the integrated probability within a given range of the peak of a galaxy's redshift posterior is above some base threshold. Defining a catastrophic outlier rate as the excess probability outside of the range $z_{low} - 0.15*(1+\bar{z})$ - $z_{high} + 0.15*(1+\bar{z})$, where $z_{low}$ and $z_{high}$ are the lower and upper bin bounds respectively and $\bar{z}$ is the average redshift of the bin, we find that probability outside of this range is $\sim3$\% for each bin which similar to that of the outlier rate estimated for the peak redshift.

The right panel in Figure~\ref{fig:photoz} shows the summed $p(z)$ distribution of the individual bins, where we have weighted each $p(z)$ in a given bin by its corresponding $P_i$ value. Values relevant to this analysis for each bin can be found in Table~ \ref{tab:mean_values}.

To increase the signal to noise of the lensing measurement, we also optimally weight each foreground galaxy by its expected lensing efficiency via
\begin{equation}
\left \langle \frac{D_l D_{ls}}{D_s} \right \rangle = \int_0^{z_{max}} p(z) \frac{D_{A}(0,z) D_{A}(z,z_{LBG})}{D_{A}(0,z_{LBG})} dz
\end{equation}
where $\langle {D_lD_{ls}}/{D_s} \rangle$ is the average, geometric efficiency, $p(z)$ is the redshift probability density for a given galaxy, and $z_{max}$ is the maximum redshift of the $p(z)$ which, given the run of BPZ, is $z_{max}=5.0$. $D_{A}(z_1,z_2)$ is the angular diameter distance from redshift $z_1$ to redshift $z_2$, and $z_{LBG}$ is the redshift of the background LBG sample which is $z\approx4$. The weight of each foreground galaxy in the correlation is then $P_i \langle {D_lD_{ls}}/{D_s} \rangle$.

\begin{figure}
\includegraphics[width=0.495\textwidth]{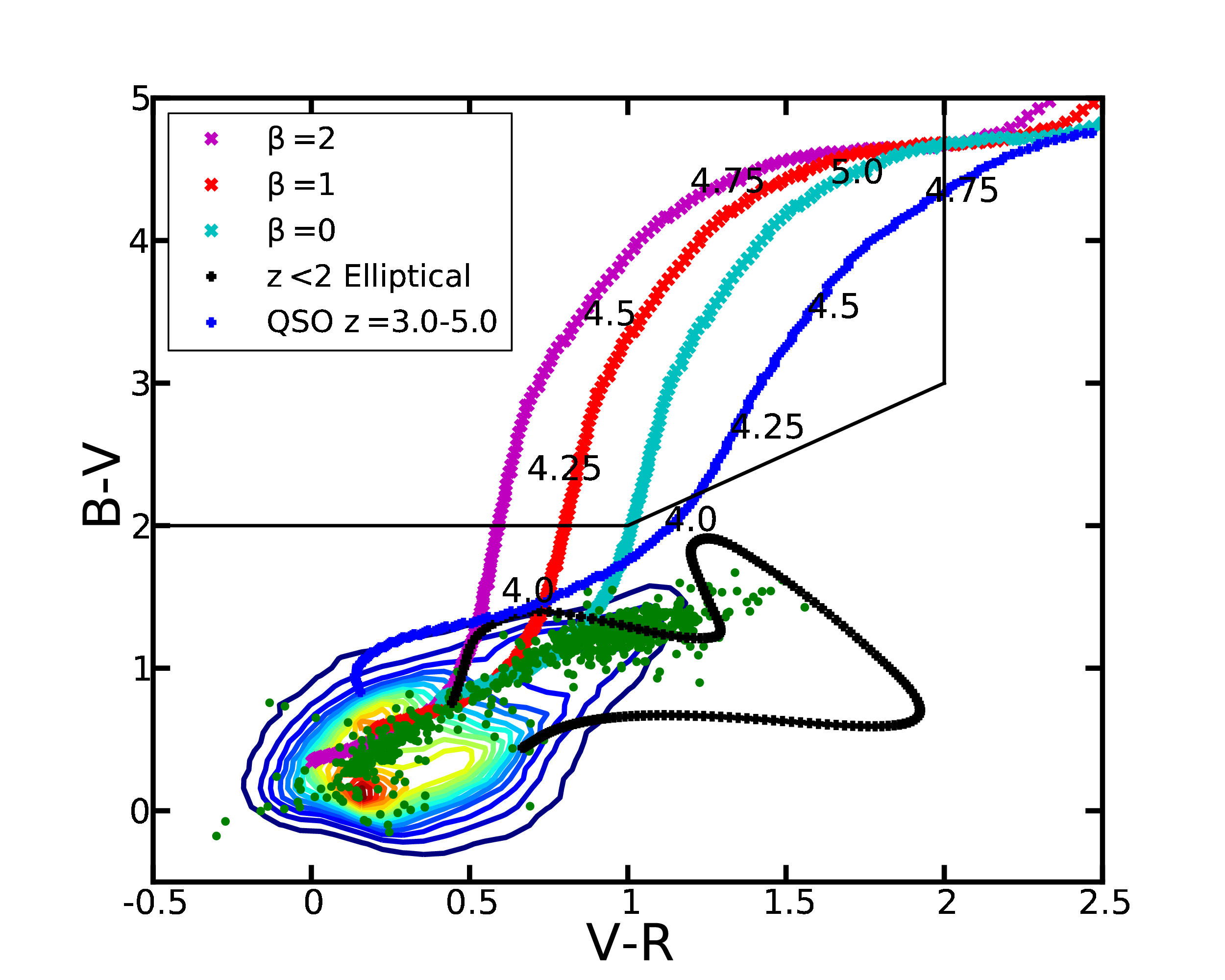}
\caption{\label{fig:lbg_selection}
Color-Color selection for redshift $z=4.0$ $B$ band dropouts. Magenta, red, and cyan $X$s are model Lyman Break Galaxies with different UV continuum slopes at redshifts indicated along the track.  The black bounding box shows the LBG selection criteria in color space. Density contours are DLS galaxies with a $B$ band $S/N>2$ detection. Green points are stars detected in DLS. The black track is the evolution of an $E_{0}$ galaxy CWW template from redshift $0<z<2$. The blue track is a sample quasar spectrum redshifted between $z=3$ and $ z=5$.}
\end{figure}

\begin{table}
  \begin{center}
    \caption{Redshift bin properties. \label{tab:mean_values}}
    \begin{tabular}{ccccc}
      \hline\hline \\
      Bin Redshift & \multicolumn{1}{c}{\# Galaxies} & 
      \multicolumn{1}{c}{$\langle z \rangle$} & 
      \multicolumn{1}{c}{$\left \langle \frac{D_l D_{ls}}{D_s} \right \rangle^{a}$} & 
      \multicolumn{1}{c}{$ \langle P \rangle $} \\ \\
      \hline \hline \\
      $0.43 < z < 0.50$ & $122\,$K & 0.49 & 636 & 0.34 \\
      $0.50 < z < 0.56$ & $114\,$K & 0.53 & 660 & 0.34 \\
      $0.56 < z < 0.63$ & $119\,$K & 0.59 & 670 & 0.35 \\
      $0.63 < z < 0.70$ & $100\,$K & 0.66 & 676 & 0.34 \\
      $0.70 < z < 0.79$ & $113\,$K & 0.76 & 672 & 0.36 \\
      $0.79 < z < 0.88$ & $114\,$K & 0.86 & 660 & 0.33 \\
      $0.88 < z < 1.00$ & $130\,$K & 0.98 & 635 & 0.33 \\
      $0.43 < z < 1.00$ & $457\,$K & 0.71 & 653 & 0.74 \\ 
      \hline
    \end{tabular}
  \end{center}
    $^a$ in Mpc/$h$
\end{table}

\subsection{Lyman Break Galaxy Selection}\label{sec:lbg_selection}

To cleanly select a sample of high redshift source galaxies and avoid contamination with the foreground lenses, we employ the Lyman Break criterion as outlined in \citet{Guhathakurta90} and \citet{Steidel99}. We select B band dropout galaxies at $z\sim4.0$ to use as lensed, source galaxies. These galaxies have well understood luminosity functions (LFs) \citep{Steidel99, Sawicki06, Bouwens07, vanderBurg10}. The selection for these galaxies in DLS is shown in Figure \ref{fig:lbg_selection}. Using the tracks in color-color space for high redshift galaxies with different UV continuum slopes, we select a region that avoids low redshift contaminants and stars. The $B-V$ cut is selected to avoid dusty red cluster galaxies that can mimic the color of LBGs while the other cuts avoid both the CWW Elliptical track as well as dwarf stars. This color-selection is
\begin{equation}
(B-V)>2 \cap (V-R)<2 \cap (B-V)>(V-R)+1
\end{equation}
for $z=4.0$ B band dropouts. To ensure that the galaxy has dropped out of the B band and is not just due to magnitude scatter or deeper/shallower data, we implement a signal to noise cut as well as a V band brightness cut that is equal to the average depth of B band.
\begin{eqnarray}
(S/N)_{B}<1 \cap (S/N)_{V}>4 \cap (S/N)_{R}>5 \nonumber \\ \cap \, m_V<25.5
\end{eqnarray}
This selection yields $\sim 12,000$ LBGs over the whole survey.

It should be stated that this selection of LBGs is degenerate with a selection of quasars (QSOs) at the same redshift as evidenced by the QSO track in Figure~\ref{fig:lbg_selection}. However, as shown by the measured QSO function within the DLS \citep{Glikman11} we expect the number counts of QSOs to be two orders of magnitude below that of the LBGs for these magnitudes and thus not a significant contaminant.

For the measurements in \S\ref{sec:optimal_results} and \S\ref{sec:tomography_results} we impose a cut of $m_R < 24.8$ on the LBGs to ensure they are above the LBG completeness limit. However, in \S\ref{sec:mag_results} we use the full range of detected LBG magnitudes to show consistency with predicted LBG LFs.

\begin{figure*}
\includegraphics[width=0.495\textwidth]{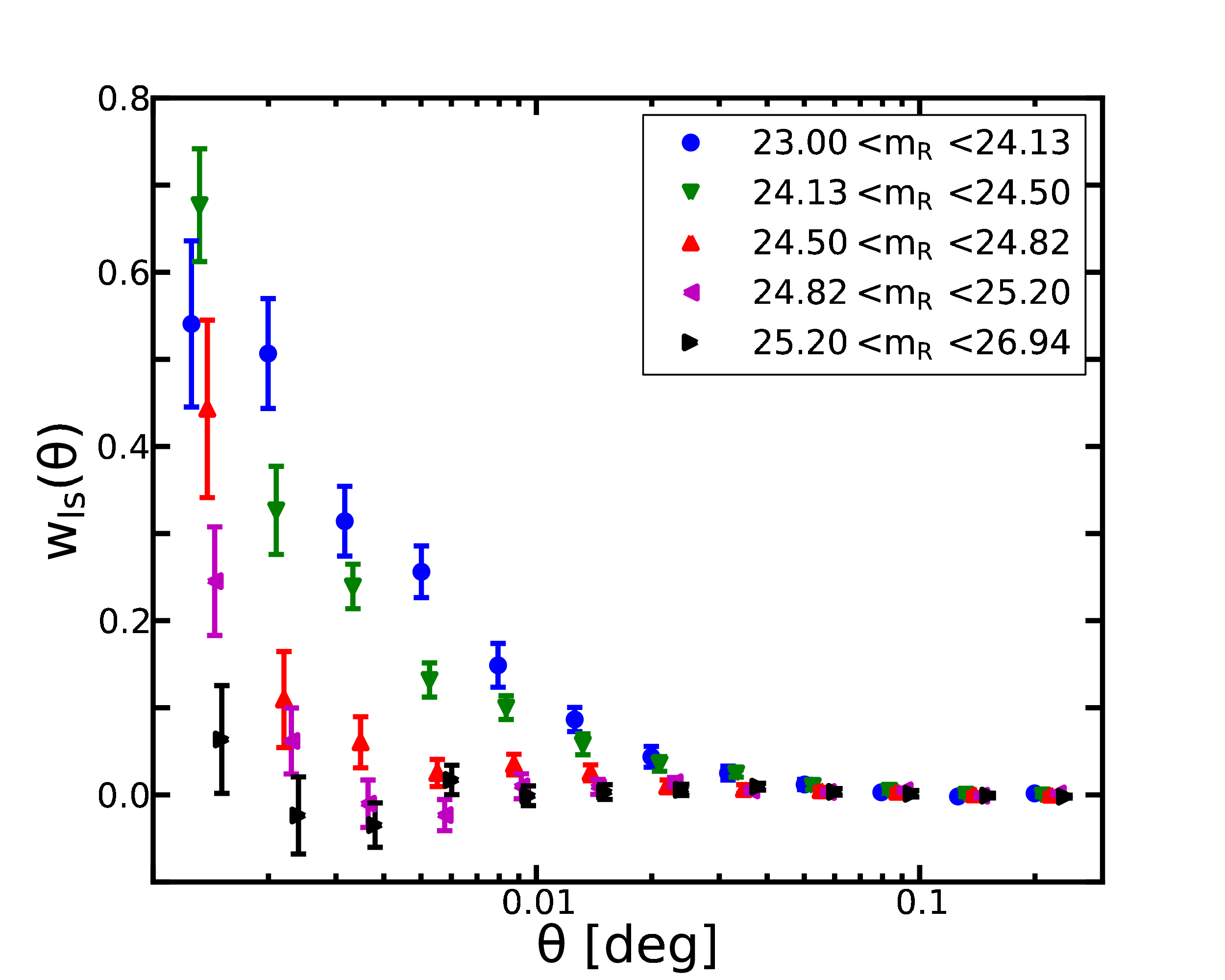}
\hfill
\includegraphics[angle=270,width=0.495\textwidth,trim=8.1in 0 0 0]{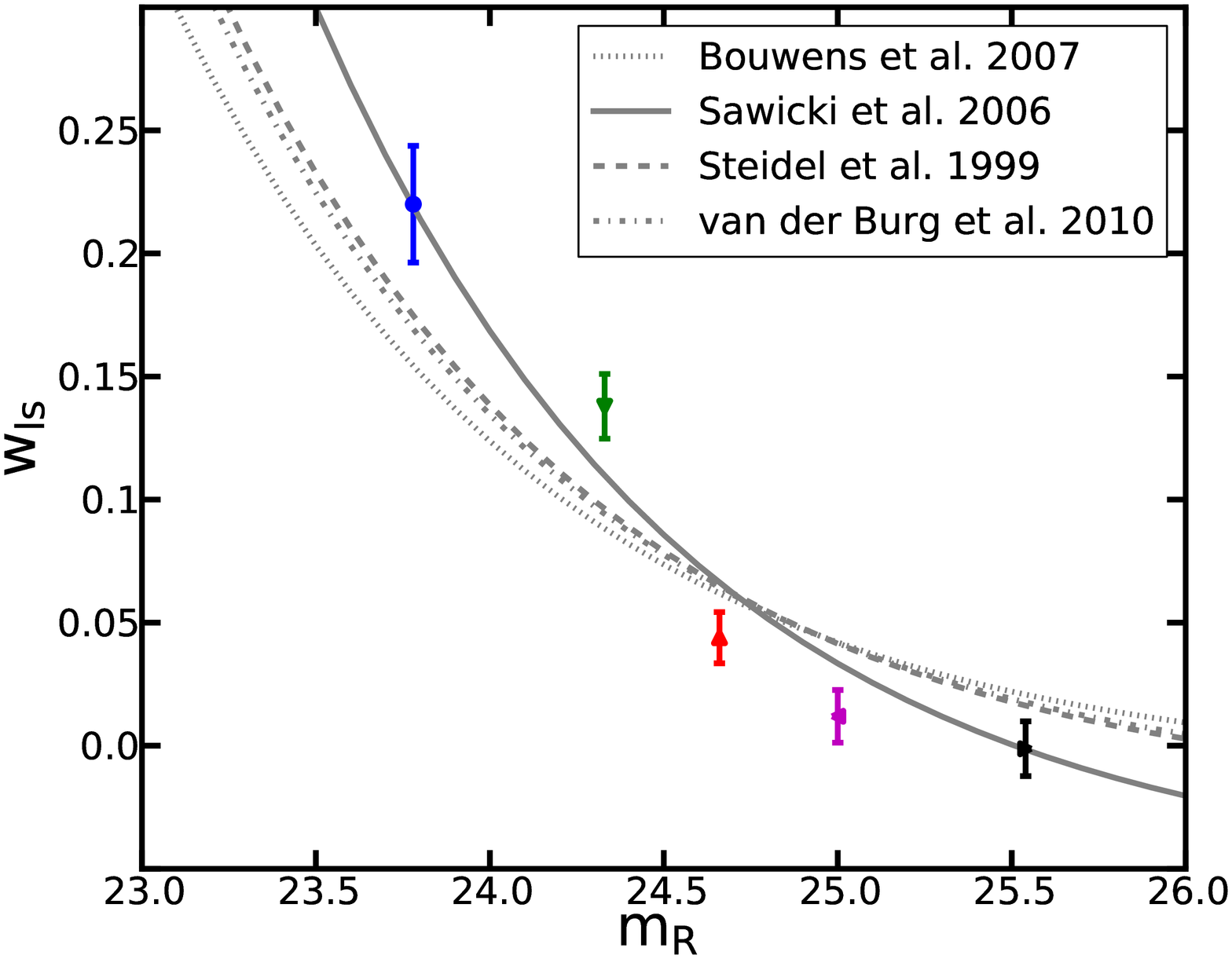}
\caption{\label{fig:mag_results}
Lens-source cross correlation due to magnification of background LGBs as a function of LBG magnitude.  The left panel shows the full angular cross-correlation for each sample against foreground galaxies with $m_R < 24$ and photometric redshift $0.4 < z_p < 1.0$.  In the right panel, we plot the lens-source cross-correlation for a single bin $0.001 < \theta < 0.01$ deg for each magnitude bin and compare it to the expected scaling for LFs from \citet{Bouwens07}, \citet{Sawicki06}, \citet{vanderBurg10}, and \citet{Steidel99}.}
\end{figure*}

\section{Results}\label{sec:results}

To estimate the magnification-induced cross-correlation we use the \citet{LS93} estimator
\begin{equation}
w_{ls}(\theta) = \frac{\langle d_l d_s \rangle - \langle d_l R_s \rangle - \langle R_l d_s \rangle + \langle R_l R_s \rangle }{\langle R_l R_s \rangle }
\end{equation}
where $\langle d_l d_s \rangle$ (here we use lower case $d$ to differentiate between this quantity and the angular diameter distance, $D$) is the number of pairs between the lensing and source galaxies in a given angular bin and $R_i$ is a random sample generated using the density and spatial extent of either the lensing or source samples.  For most of the calculations, the data and random points are weighted by some factor $w_i$ making $\langle d_l d_s \rangle$ the sum of the products of the weights for each pair with the same set of weights are applied to the random realizations.

Starting with the basic DLS survey footprint, we further mask out regions with bad seeing, high dust extinction and shallow depth to ensure a spatially uniform selection of foreground and background objects (see Appendix~\ref{sec:systematics} for details). This reduces the effective area of the DLS from 20 to 13.5 $\rm{deg}^{2}$.  We also measure the magnification in each of the 45 subfields individually rather considering the survey as a whole to minimize the effects of varying depth between the subfields. This yields a natural, consistent angular scale for assembling jack-knife subsamples for the purposes of calculating measurement errors. 

\subsection{Lyman Break Galaxy Magnitude Bins}\label{sec:mag_results}

We measure the foreground galaxy - LBG angular angular cross-correlation as a function of LBG magnitude. As shown in Eq.~\ref{eq:mag_correlation}, the magnification signal is expected to scale as the slope of the LF: $w_{ls} \sim \alpha(m) - 1$. We split the LBG sample into 5 magnitude bins, spanning the range $23 < m_R < 27$ and sampling different parts of the LF with roughly equal numbers of LBGs in each bin.  The measured angular correlations $w_{ls}(m)$ are shown in the left panel of Figure~\ref{fig:mag_results}. The observed density excess of LBGs around foreground galaxies is seen to decrease for fainter galaxies, which correspond to the shallowed part of the luminosity function.  To explore this more quantitatively, we consider a single angular bin running from $0.001 < \theta < 0.01$ deg for each of the samples and we show the corresponding LBG density change in the right panel of Figure~\ref{fig:mag_results}. We can compare these values, up to a multiplicative scaling factor, with estimates of the slope of the LBG luminosity function from \citet{Bouwens07}, \citet{Sawicki06}, \citet{vanderBurg10} and \citet{Steidel99}. As can be seen in the figure, the magnitude dependence of the magnification signal is properly recovered. 

\begin{figure}
\includegraphics[angle=270,width=0.495\textwidth]{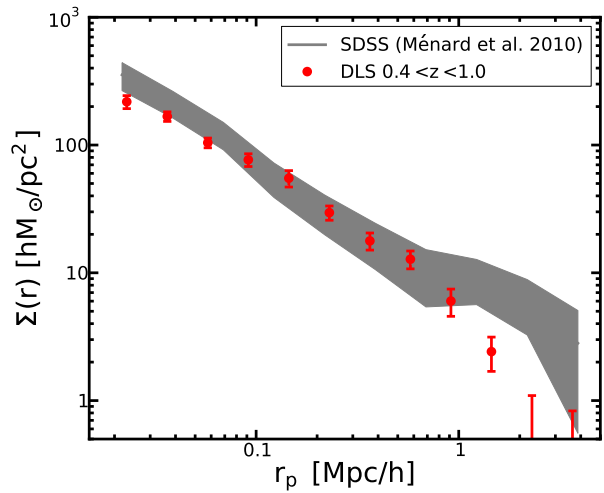}
\caption{\label{fig:mag_optimal} 
Optimally-weighted magnification-reconstructed mass density $\Sigma$ as a function of projected radius $r_p$, using all LBG candidates from $23 < m_R < 24.8$ and foreground lenses between $0.4 < z_p < 1.0$.  This mass profile is consistent with the similarly derived profile from SDSS-based magnification in \citet{Menard10}.}
\end{figure}

\begin{figure*}
\includegraphics[width=0.495\textwidth]{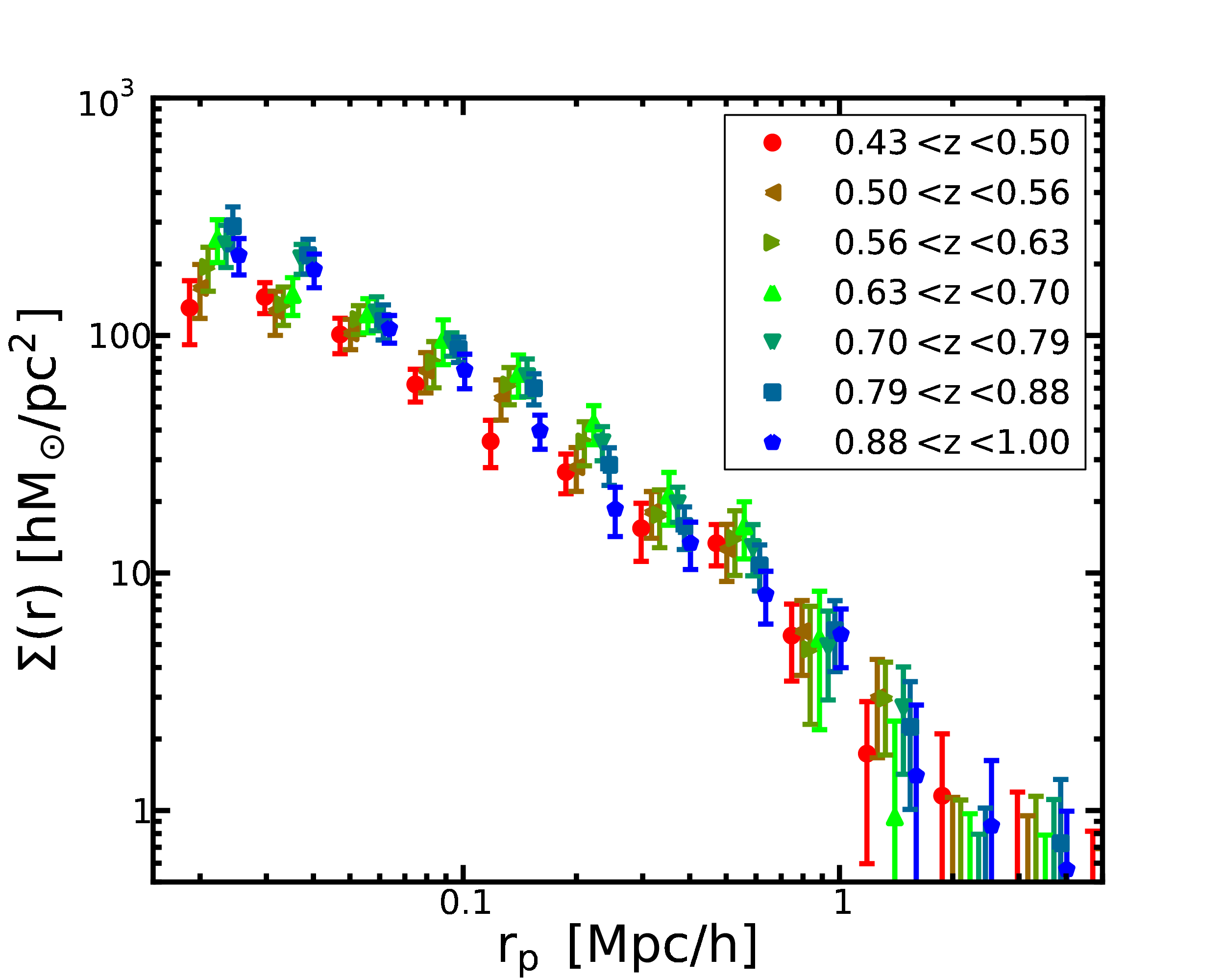}
\includegraphics[width=0.495\textwidth]{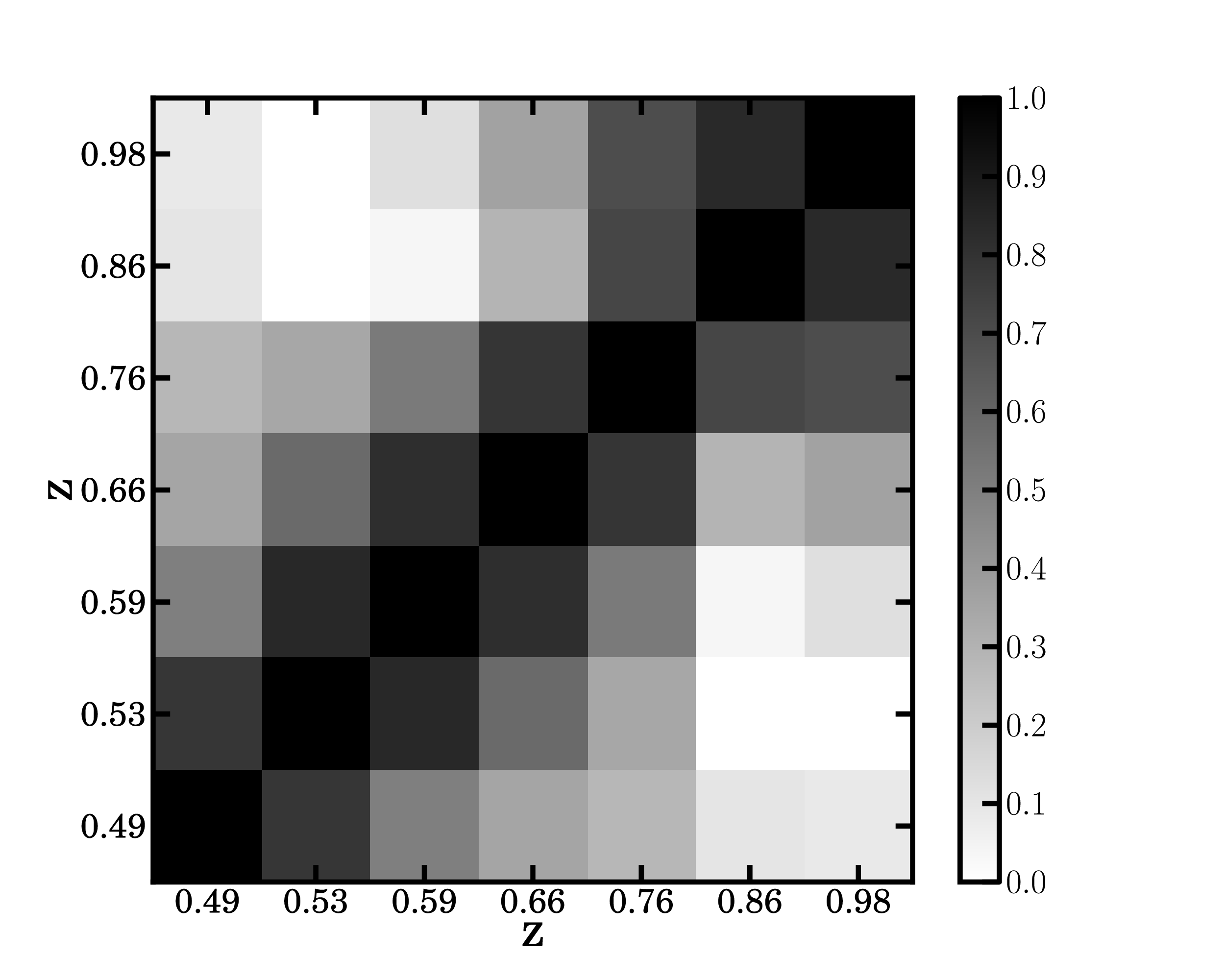}
\caption{\label{fig:tomo_corr}
The left panel shows the mass profile reconstruction for each of the 7 photometric redshift bins.  For each bin, we calculate the tomographic ratio $\hat{\mathcal{R}}(z)$  over the range $0.4 < r_p < 1.25$ Mpc/$h$ as given in Equation~\ref{eq:bias_ratio}.  The right panel shows the covariance matrix for these 7 bins.  Since $\hat{\mathcal{R}}(z)$ is a function of both the lens-source cross-correlation and the lens autocorrelation, any correlation between bins in $\hat{\mathcal{R}}(z)$ is a strong indicator of the degree of overlap between redshift bins. We find good photometric redshift bin segregation, as required for tomography. }
\end{figure*}

\subsection{Galaxy-Mass Correlation}\label{sec:optimal_results}

In order to combine the measurements from different magnitude bins and maximize the overall S/N of the magnification measurement, we apply the optimal estimator described in \citet{Menard02}, weighting each LBG by the corresponding $\alpha(m) - 1$ value from the \citet{Sawicki06} LF, which best matched the observed scaling with LBG magnitude from Figure~\ref{fig:mag_results}. Other choices of luminosity function for the optimal estimator give consistent results within the error bars of the measurements, however, they do tend to bias the measurements low or high depending on the relative value of $\langle \alpha - 1 \rangle$ for the LF. From these measurements we can infer the mean surface mass density around the foreground galaxies and compare the results to those obtained by \citet{Menard10} who used magnification measurements of quasars in the SDSS. To convert the measured density change into a surface mass density, we first compute the average $\Sigma_{crit}$ for each lens sample based on the average lensing kernel for each of the $N$ galaxies in a given redshift bin:
\begin{equation}
\langle \Sigma_{crit} \rangle = \frac{c^2}{4\pi G} \left (N^{-1} \sum_i \left \langle \frac{D_l D_{ls}}{D_s} \right \rangle_i \right )^{-1}.
\end{equation}
With this in hand the conversion from magnification angular correlation to surface mass density is
\begin{equation}
\Sigma(r_p) = \frac{w_{ls}(r_p)}{2}\langle \Sigma_{crit} \rangle .
\end{equation}

For the LBG sample of $m_R < 24.8$ and all foreground galaxies between $0.4 < z < 1.0$, we find a detection S/N of $\sim 20$, as seen in Figure~\ref{fig:mag_optimal}.  For comparison, we also show the results from the mass profile found for SDSS galaxies via magnification from \citet{Menard10}. The DLS measurements are at higher redshift ($z \sim 0.4$ for the SDSS sample) and redshift-selected rather than magnitude-selected, so some level of disagreement is expected.  However, from a qualitative standpoint, there is good consistency between the two measurements over a large range in scale.

\begin{figure*}
\includegraphics[angle=270,width=0.495\textwidth,trim=8.1in 0 0 0]{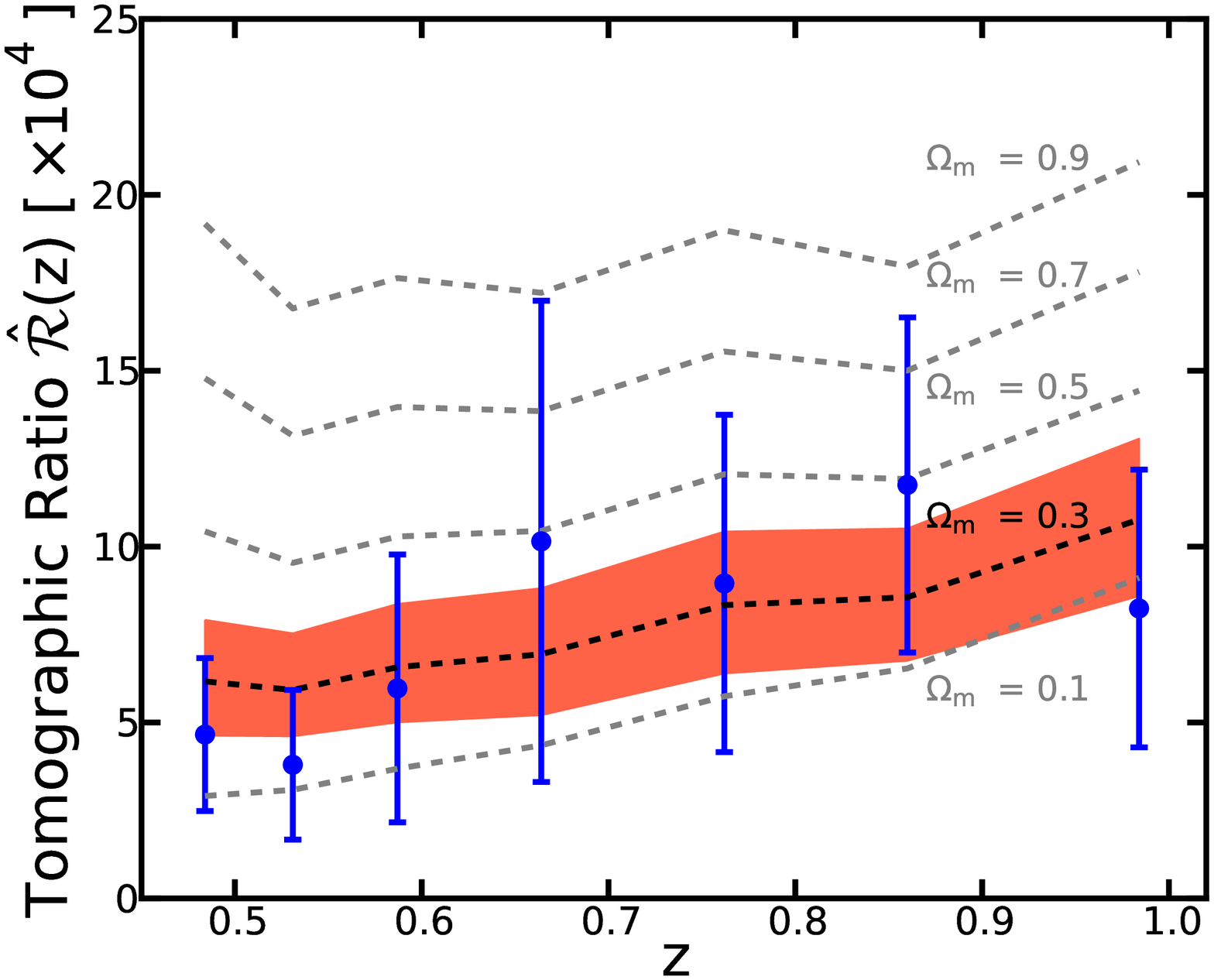}
\includegraphics[width=0.495\textwidth]{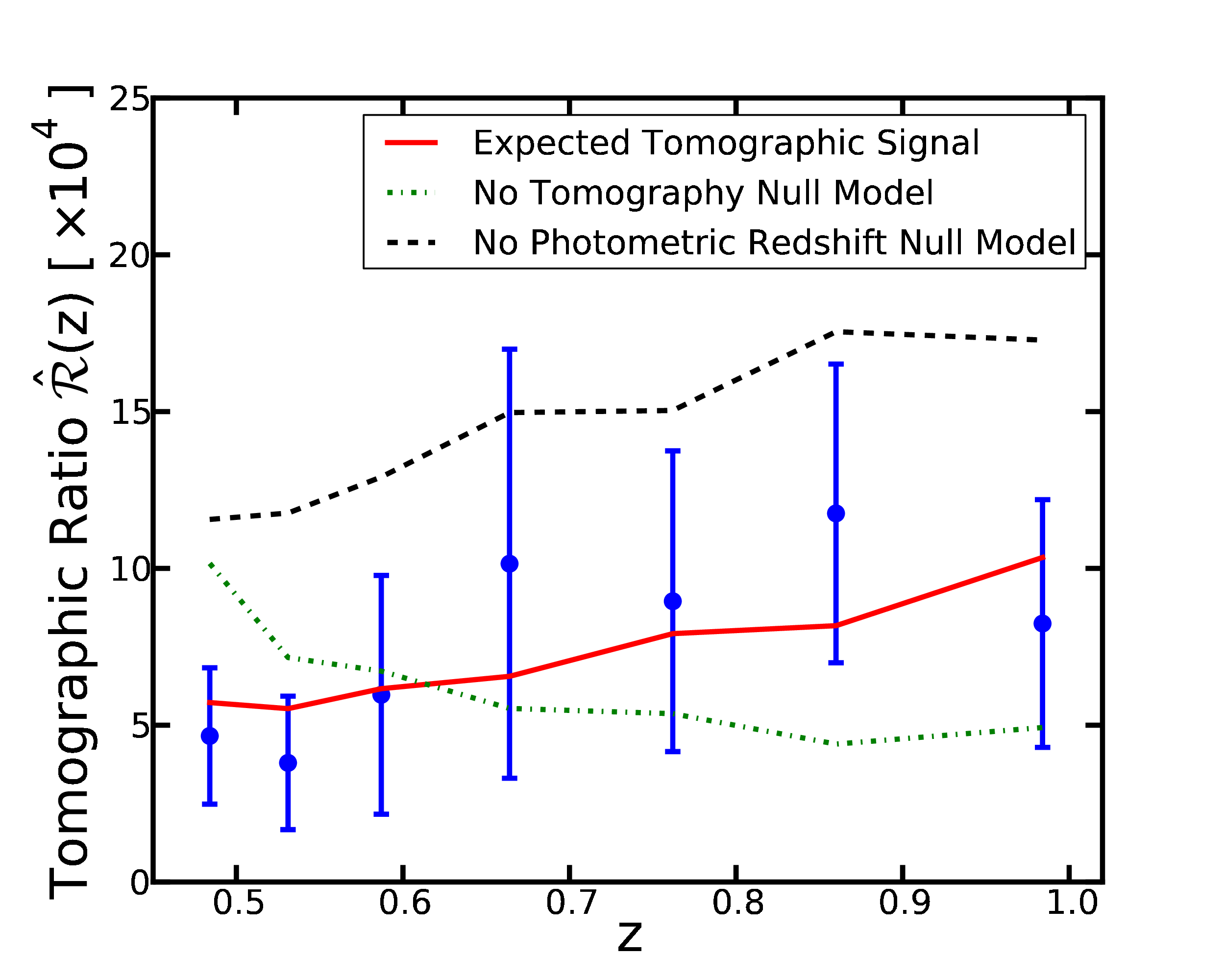}
\caption{\label{fig:tomo_results} 
Tomographic ratio ($\hat{\mathcal{R}}(z)$) for photometric redshift bins over $0.4 < z < 1.0$ on scales $0.4 < r_p < 1.25$ Mpc/$h$.  The left panel shows $\hat{\mathcal{R}}(z)$ for cosmologies over a range of $\Omega_m$ values. The shaded region shows a variation of $\pm$0.1 with respect to the fiducial value of $\sigma_8$. For fixed $\sigma_8$, we reject a matter dominated universe at $>7.5 \sigma$ and prefer a universe with $\Omega_m =0.3$.  The right panel compares the measured $\hat{\mathcal{R}}(z)$ with two potential null models.  We reject the scrambled photometric redshift model at $3.8\sigma$, however the measurements disfavor the no-tomography theoretical null model at only $2.6\sigma$.}
\end{figure*}

\subsection{Tomography}\label{sec:tomography_results}

We now explore the redshift-dependence of the lensing signal. We estimate the galaxy-mass correlation function for 7 subsamples of foreground galaxies with varying redshifts as described in \S\ref{sec:lens_selection}. The left panel of Figure~\ref{fig:tomo_corr} shows the values of $\Sigma(r)$ as a function of lens redshift. The observed redshift trend is due to a number of effects: due to our brightness and color selection, galaxies with different types and masses are selected in different redshift bins. In addition the shape of the redshift distribution differs from a bin to another one.

As mentioned previously, we can combine the galaxy-mass cross-correlation and galaxy auto-correlations through the quantity $\hat{\mathcal{R}}(z)$
introduced in \S\ref{sec:tomographic_theory}
to obtain a quantity which, on large scales, does not depend on the galaxy properties.
To measure $\hat{\mathcal{R}}(z)$, we use a single bin of constant physical radius over $0.4 < r_p < 1.25$ Mpc/$h$ for each photometric redshift bin. This choice of spatial scales is limited by the range over which $\hat{\mathcal{R}}(z)$ is independent of galaxy bias (see Appendix \ref{sec:hod_theory}), as well as the acceptable scales where $w_{ls}$ and $w_{ll}$ have good S/N. 

Before discussing the $\hat{\mathcal{R}}(z)$ measurements themselves, we consider the normalized correlation matrix for $\hat{\mathcal{R}}(z)$ in the right panel of Figure~\ref{fig:tomo_corr}.  Since the covariance between redshift bins in $\hat{\mathcal{R}}(z)$ is a combination of the covariance for both the magnification and foreground autocorrelation signals, the level of correlation between redshift bin pairs is expected to be sensitive to the redshift overlap between those bins. The correlation which we find is consistent with what we expect from the distributions in Figure~\ref{fig:photoz}.  This suggests that we have the level of photometric redshift segregation required to observe tomographic lensing.  We also find that this is robust under a variety of cuts in $P$-threshold (see \S\ref{sec:lens_selection}) and photometric and radial binning. In using this matrix for deriving confidences, we correct the inverse covariance as described in \citet{Hartlap07}.

We present measurements of $\hat{\mathcal{R}}(z)$ in Figure~\ref{fig:tomo_results} for each of the 7 photometric redshift bins, along with a number of theoretical models.  The left panel compares the observed $\hat{\mathcal{R}}(z)$ with several flat cosmologies with increasing $\Omega_m$ and fixed $\sigma_8$ and $h$.  For a fixed set of redshift bins, the amplitude of $\hat{\mathcal{R}}(z)$ scales quite strongly with $\Omega_m$ since the overall amplitude of the lensing signal is directly proportional to the mean matter density. We also plot a shaded region around $\Omega_m=0.3$, varying $\sigma_8$ by $\pm 0.1$ around the feducial value of 0.8 to give the reader a sense of the usual $\sigma_8$, $\Omega_m$ degeneracy. We find that the measurements reject a flat, matter dominated model at $>7.5\sigma$, preferring a cosmology with $\Omega_m \sim 0.3$ ($\chi^2/ \nu = 1.7$) for a fixed value of $\sigma_8$. 

However, this consistency does not necessarily verify that we are observing tomographic lensing. To do so, we consider two potential null models, one observational and one theoretical.  The observational null model assumes that the photometric redshift binning has failed completely and the redshift bins are essentially random subsamples of the full foreground lens sample.  The theoretical model assumes that $K(\chi)$ from Equation~\ref{eq:mag_kernel} has been replaced by a constant, i.e. that the lensing kernel is flat as a function of redshift. There is no obvious value for us to use; therefore we set this constant to the mean kernel value of the lensing sample.

We plot the results of these tomographic null tests in the right panel of Figure~\ref{fig:tomo_results}.  The measurements reject the scrambled lens redshift null test at $3.8\sigma$ and prefer a $\Lambda$CDM universe at $4.8 \sigma$, as one would expect based on the correlation matrix in Figure~\ref{fig:tomo_corr}.  We reject the fixed constant theoretical model at $2.6\sigma$ and prefer the concordance cosmology to this model at $3.0 \sigma$.  This mild rejection of the theoretical null model is not entirely surprising given that the lensing kernel across these samples is relatively flat, as seen in Table~\ref{tab:mean_values}.  Had we been able to make measurements at either higher or lower redshift, rejecting the null theoretical model might have been possible.  We explored pushing beyond the lens redshift limits shown in the right panel of Figure~\ref{fig:photoz}, but the combination of small survey angular extent and photometric redshift degeneracies made measurements at those redshifts unreliable.

\section{Discussion and Conclusions}\label{sec:conclusions}

Using data from the Deep Lens Survey, we have shown a robust, $S/N > 20$ detection of the cosmic magnification of background LBGs by foreground galaxies, producing a halo mass profile similar to that observed in \citet{Menard10}.  In addition, we used photometric redshifts to divide the foreground sample into 7 redshift-selected bins, recovering a strong detection in each as well as demonstrating that the measurements had the expected level of redshift segregation. By combining these lensing measurements with the measured angular autocorrelation in each foreground sample, we effectively de-biased the measurement, isolating the cosmological signal $\hat{\mathcal{R}}(z)$.

We tested these measurements of $\hat{\mathcal{R}}$ in two ways.  First, we compared it to the expected signal for flat universes with fixed $h$ and $\sigma_8$ and varying $\Omega_m$.  In this test, we found that the results were consistent with the concordance cosmology and rejected a flat matter dominated universe at $>7.5\sigma$.  Second, we tested the tomographic nature of the $\hat{\mathcal{R}}(z)$ measurement by considering two null models: an observational null model where we assumed that, despite indications otherwise, the photometric redshifts failed to separate the foreground samples; and a theoretical null model where we assumed that the lensing kernel was a constant set to the mean value of the kernel for this sample.  We reject the observational null model at $3.8\sigma$ but the theoretical null model was only rejected by the data at $2.6\sigma$, owing to the fact that the lensing kernel only varied by $\sim 10\%$ over the 7 redshift bins and higher and lower redshift samples were unreliable due to the limitations of the survey.

From these results we can draw several conclusions: First, that magnification can yield high signal to noise without the necessity for complex shape measurements; second, that magnification can be combined with autocorrelation to suppress galaxy bias and constrain the integrated dark matter distribution. While the method in this paper obviates the need for it, fitting a full halo model to the data is feasible and would allow this measurement to both constrain cosmology directly as well as measure the galaxy bias. Upcoming papers will address this extension.

Future work will include combining with measurements of galaxy-galaxy lensing measured using the same foreground lens sample. This will leverage the additional constraining power of both the galaxy bias and the multi-source and multi-lens weak lens tomography \citep{VanWaerbeke10}. With the large range of redshifts as well as the ability of the magnification to utilize unresolved objects, we can probe mass-luminosity relations at $z\sim1.0$ as well as the dust contained in high redshift galaxies \citep{Menard10}.

\section{Acknowledgements}

We thank Paul Thorman for discussions, Jim Bosch and Martin Dubcovsky for their coding advice, Irina Udaltsova for help computing significances, and Perry Gee for help with the DLS photometry and database. We also thank the reviewer for their comments. This work was supported by NSF Grant AST-1009514. Brice M\'{e}nard is supported by the NSF and the Alfred P. Sloan foundation. Ami Choi acknowledges support from the European Research Council under the EC FP7 grant number 240185 and NSF Grant AST-1108893.

Funding for the Deep Lens Survey has been provided by Bell Labs Lucent Technologies and NSF grants AST 04-41072 and AST 01-34753. Observations were obtained at Cerro Tololo Inter-American Observatory and Kitt Peak National Observatory. CTIO and KPNO are divisions of the National Optical Astronomy Observatory (NOAO), which is operated by the Association of Universities for Research in Astronomy, Inc., under cooperative agreement with the National Science Foundation.

We thank the PRIMUS team for sharing their redshift catalog. Funding for PRIMUS has been provided by NSF grants AST-0607701, 0908246, 0908442, 0908354, and NASA grant 08-ADP08-0019. This paper includes data gathered with the 6.5 meter Magellan Telescopes located at Las Campanas Observatory, Chile.

\appendix

\section{Tomographic Ratio Bias Dependence}\label{sec:hod_theory}

\begin{table*}
\begin{center}
\caption{\label{tab:hod_models} Mean Redshift and HOD Bias}
\begin{tabular}{ccrrr}
\hline\hline
 & HOD Model Name & $\langle z \rangle$ & $M_0$ & $b_g$ \\
\hline
\multirow{3}{*}{DLS G-G Lensing} & DLS A & 0.58 & 0.13 & 1.60 \\
 & DLS B & 0.58 & 0.02 & 1.43 \\
 & DLS C & 0.53 & 0.01 & 1.27 \\
 \hline
\end{tabular} \\
\begin{tabular}{ccrrrr}
\hline\hline%
& HOD Model Name & $\langle z \rangle$ & $M_{min}$ & $M_1'$ & $b_g$ \\
\hline
\multirow{3}{*}{\citet{Wake11}$^a$} & Wake A & 1.1 & 0.31 & 1.45 & 2.59 \\
 & Wake B & 1.1 & 0.17 & 0.76 & 2.37 \\
 & Wake C & 1.1 & 0.10 & 0.42 & 2.17 \\
\hline
\end{tabular} \\
\end{center}
All Masses are in $10^{13}$ $M_{\odot}/h$. \\
$^a$Several HOD parameters are fixed for this measurement, they are: $M_0~=~M_{min}$; $\sigma_{\rm{log}M}~=~0.15$; $\alpha~=~1.0$.
\end{table*} 

To estimate the bias dependence of $\mathcal{R}$ we implement a halo model power spectrum code \citep{Seljak00} with a \citet{ShethTormen02} mass function. Within this model, we use two different halo occupation distributions (HODs), based on forms given by \citet{Mandelbaum05} and \citet{Zheng07} (M05 and Z07, hereafter, respectively). The galaxy bias for the magnification and autocorrelation are determined by the first ($\langle N \rangle$) and second moments ($\langle N(N-1) \rangle$) of the HOD respectively. So long as the ratio between these two moments is Poissonian ($<N>^2 \sim <N(N-1)>$), then any scale dependence in the bias for the samples will cancel out, leaving Equation~\ref{eq:bias_ratio} bias independent.

The M05 HOD breaks the expected number of galaxies per halo into satellite ($\langle N_s | M \rangle$) and central ($\langle N_c | M \rangle$) galaxies. A halo has a single central galaxy if the mass of the halo, $M$, is above some threshold mass $M_0$. For the satellite galaxies, the expected number of galaxies in a halo of mass $M$ is
\begin{eqnarray}
\langle N_s | M \rangle \propto
   \begin{cases}
  M^2 & \text{if } M < 3M_0 \\
  M & \text{if } M > 3M_0  
  \end{cases}
\end{eqnarray}
The average number of total galaxies in a halo is then
\begin{equation}
\langle N | M \rangle = \langle N_c | M \rangle + \langle N_s | M \rangle
\end{equation}
For values of $M_0$ we use the measured values in three different luminosity samples at $z\sim0.5$ as observed in the DLS by galaxy-galaxy lensing (private communication A. Choi).

The Z07 HOD also uses a central/satellite galaxy model as a function of halo mass. The functional form of the central term is
\begin{equation}
\langle N_c | M \rangle = \frac{1}{2}\left[ 1+\rm{erf} \left(\frac{\rm{log}(M)-log(M_{\rm{min}})}{\sigma_{\rm{log}M}}  \right )\right ]
\end{equation}
where $M_{\rm{min}}$ is the minimum mass for a halo to have one galaxy and $\sigma_{\rm{logM}}$ is the width of the central galaxy turn on. The satellite galaxy term is
\begin{equation}
\langle N_s  | M\rangle= \left(\frac{M-M_0}{M_1'}\right)^\alpha
\end{equation}
where $M_0$ is the minimum mass for a halo to host satellite galaxies (note this is distinct from the $M_0$ in the Mandelbaum model), and $M_1'$ is the mass differential at which a halo is expected to have one satellite galaxy. The average number of galaxies occupying a halo of a given mass is then
\begin{equation}
\langle N |M \rangle = \langle N_c | M \rangle (1+\langle N_s  | M \rangle)
\end{equation}
We use the measured parameters for this HOD model from \citet{Wake11} for galaxies at $z=1.1$ in three bins of stellar mass.

The form of the HOD second moment we use is from \citet{Zheng05},
\begin{equation}
\langle N(N-1) | M \rangle = 2 \langle N_s | M \rangle + \langle N_s | M \rangle ^2 
\end{equation}
where $N_s$ is the expected number of satellite galaxies in a halo of mass $M$.

Table~\ref{tab:hod_models} presents the parameters for the HODs used, along with the mean redshift and linear galaxy bias. As discussed in \S\ref{sec:data}, the redshifts for the lensing samples are bracketed by the galaxies used in the DLS galaxy-galaxy lensing analysis \citet{Choi12} and \citet{Wake11}, so we expect the results to be bounded by these models. Since the galaxies are selected by photometric redshift within fixed apparent magnitude bounds, we expect them to be slightly biased relative to field galaxies, with the bias increasing as we go to higher redshifts.  As shown in Figure~\ref{fig:hod_compare}, the various HOD models track the overall shape of the dark matter-only model reasonably well and are within $\sim 20\%$ of the dark matter ratio for all of the redshift bins even at scales as small as 0.5 Mpc/$h$. This suggests that Equation \ref{eq:bias_ratio} is adequately galaxy bias independent on these small scales and, to within the precision of this measurement, probes the magnification due to dark matter directly.

\section{Deep Lens Survey Calibration \& Photometric Redshifts}\label{sec:calibration}

\subsection{Photometric Calibration}

The DLS image processing and photometry are described in detail in an upcoming data release paper (Wittman et al. in prep.); here, we give a brief overview.  The images were de-biased, flat-fielded, and sky-subtracted using the procedures described in \citet{Wittman06}. Next, we used the global, linear least-squares algorithm known as {\it ubercal} \citep{Padmanabhan08, Wittman11} to correct for residual flat-fielding errors, after which the corrected images were cross-registered and photometrically stacked. Object detection was done with the $R$ band (the deepest and highest-resolution filter) using {\it SExtractor} \citep{Bertin96}, with magnitudes measured in the other bands regardless of S/N in that band.  We used a modified version of {\it ColorPro} \citep{Coe06} to provide robustness against variations in seeing because photometric redshifts require matched aperture photometry.  To measure a given color (e.g. $B-R$), {\it ColorPro} convolves the $R$ band image to match the $B$ image and then uses matched apertures on the seeing-matched images.  The resulting photometry showed spatial variations of up to $\sim 0.04$ mag in $z$ band and $\sim 0.03$ mag in $B$ and $V$ bands, based on the stellar locus in color-color space.  We corrected for this by applying zero-point shifts to each subfield to make the stellar locus consistent across subfields and across fields.  This was generally a straightforward correction, but the varying shape of the stellar loci with galactic latitude and longitude complicated the overall shifts of fields F1 and F2 with respect to each other and the rest of the survey.  To determine these shifts, we used overlapping parts of the Sloan Digital Sky Survey \citep{SDSSDR8} as the basis for common calibration and removed the resulting shifts, which ranged from 0.00 to 0.05 mag depending on field and filter.  We estimate that the remaining spatial variation in the photometry is at the level of 0.02 mag or less.

\subsection{Photometric Redshifts}\label{sec:photoz}

The photometric redshifts are based on the $BVRz$ photometry obtained from the {\it ColorPro} and {\it BPZ} \citep{Bpz00} software packages.  We replaced the standard templates with a set optimized in a method similar to that described in \citet{Ilbert06}.  By using spectroscopic samples from the SHELS survey (private communication M. Geller) and the PRIMUS survey (\citet{Coil11}, Cool et al. in prep.) that overlap the DLS footprint, we divided the galaxies into six galaxy types ({\it Elliptical}, {\it Sbc}, {\it Scd}, etc.) and then adjusted the SED templates to match the median rest-frame fluxes observed in the DLS photometric data as a function of wavelength. This procedure matched the colors of galaxies to the observed data, reducing ``template mismatch'' bias and variance.

We also employed a modified version of the type-redshift prior used in {\it BPZ}.  Beginning with the spectroscopic data from SHELS, we fit the prior to the observed $P(z|T,m)$ distribution using the form from \citet{Bpz00}.  This prior was extended to fainter magnitudes using the VVDS spectroscopic sample \citep{LeFevre05}.  A more detailed description of the DLS photometry and photometric redshifts will be given in Thorman \& Schmidt (in prep.).

\section{Sources of Systematic Errors}\label{sec:systematics}

\subsection{Observational Contaminants}

\begin{figure*}
\includegraphics[width=0.495\textwidth]{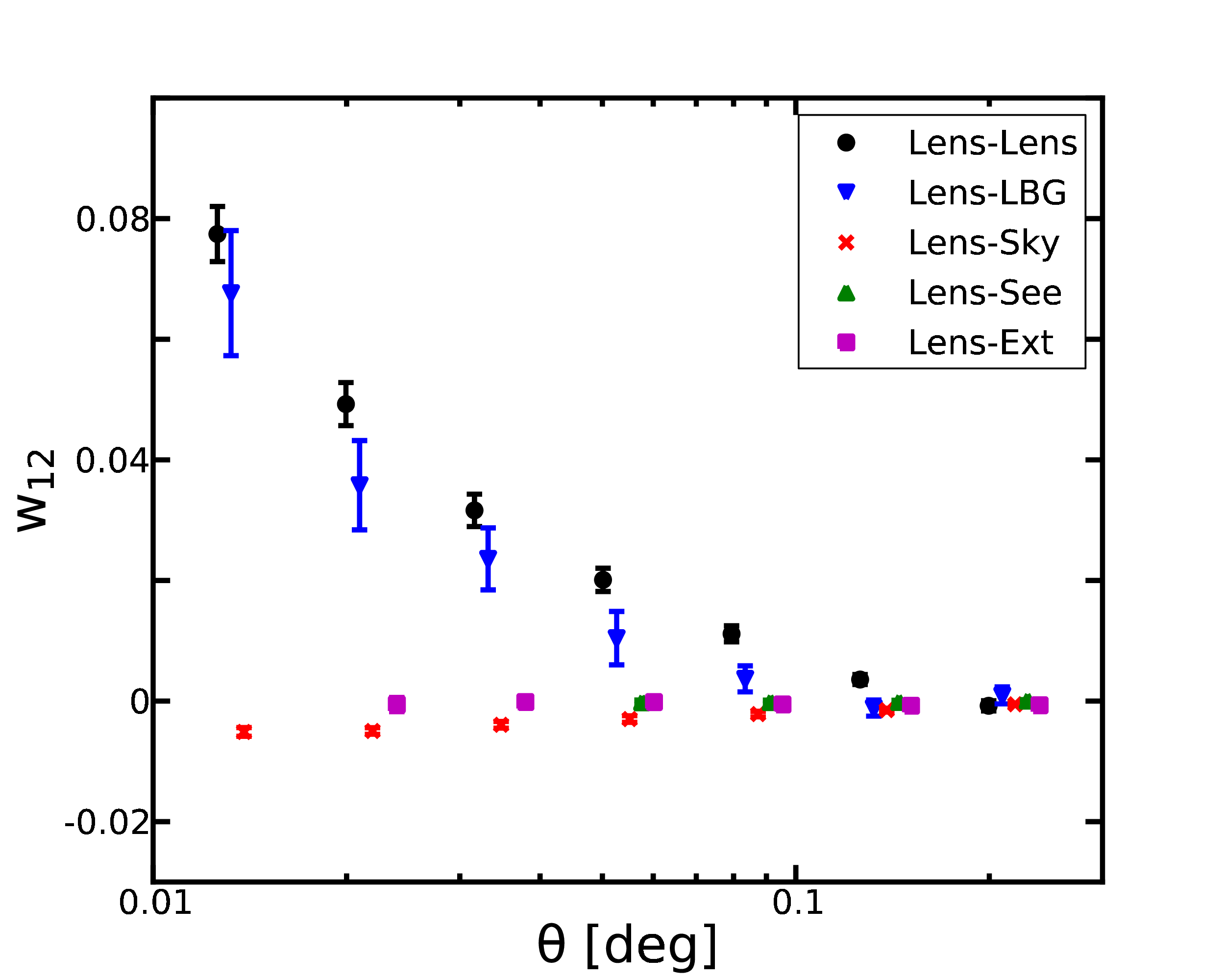}
\hfill
\includegraphics[width=0.495\textwidth]{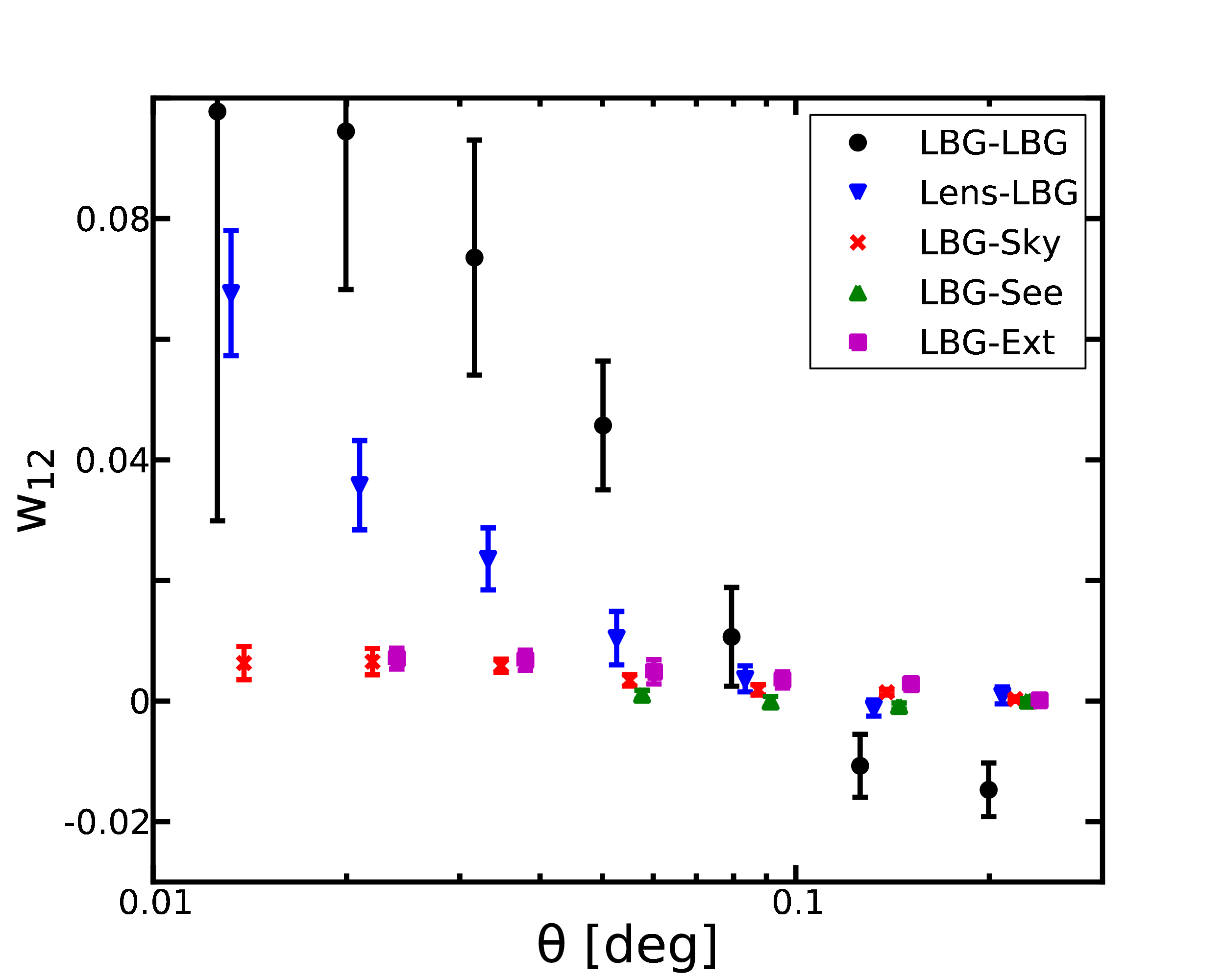}
\caption{\label{fig:DLS_sys}
Angular correlations between the foreground lens ({\it left}) and LBG ({\it right}) samples against various potential sources of systematic errors, the RMS sky noise (Sky), the mean seeing (See) full width half maximum, and extinction (Ext) from galactic dust \citep{Schlegel98}. We compare these systematics with the respective sample autocorrelation functions and the magnification signal.}
\end{figure*}

Since both the foreground and background samples are drawn from the same underlying imagery, inhomogeneities in the detection and selection process due to variations in the seeing, RMS sky noise, and extinction can induce correlations between the two samples which are unrelated to the lensing magnification signal we are trying to detect.  To reduce the effect of these biases, we exclude regions around bright stars and CCD bleeds. We also mask out the regions at the edges of each subfield where the number of stacked exposures can be as low as $\sim 4$, compared to $\sim 20$ at the center of a subfield.  This brings the nominal area down from a full survey area of 20 deg$^2$ to 13.5 deg$^2$.

To estimate the contribution of the various sources of systematic error to the angular correlations, we cross-correlate both the foreground lenses and background LBG against the seeing, RMS sky noise, and extinction for each of the 4 bands. These correlations are plotted in Figure~\ref{fig:DLS_sys} for the $R$ band. Other bands have similar values for the correlation and are not shown here.  Because we undertake a cross-correlation between a background and foreground sample, we can tolerate a certain degree of systematic correlation in one sample, so long as the other is uncontaminated. From the cross-correlations plotted in Figure~\ref{fig:DLS_sys}, we see that this is the case for the foreground lenses against the systematic sources as they do not show spatial structure, are generally consistent with zero, and are small compared to the amplitude of the magnification and autocorrelations. Since the foreground systematics are uncorrelated and the LBG systematic cross-correlations are small, the effect of these systematic errors on these measurements is small.

\subsection{Redshift Contaminants}\label{sec:red_recovery}

As stated in \S\ref{sec:data}, the most dangerous contaminant in the LBG sample are low redshift, dusty, red galaxies. If the LBG sample is contaminated by these low redshift galaxies at even 10\%, the signal could be due to physical clustering and not magnification. To test for contamination at low redshift we cross-correlate the LBG sample against the PRIMUS (\citet{Coil11}, Cool et al. in prep.) spectroscopic sample that overlaps the DLS footprint, in a single, large radial bin (100 kpc to 1 Mpc). The PRIMUS sample has $\sim8000$ spectra with roughly $\sim 2500$ LBGs in the intersection of the PRIMUS footprint with the 13.5 $deg^2$ used in this analysis. This method is similar to that of \citet{Newman08} and \citet{Matthews12} and exploits the physical clustering of galaxies to recover redshift distributions. While this spectroscopic sample does not allow us to recover the redshift distributions beyond $z\sim1$, we can utilize it as a null test for contamination from low redshift galaxies. (see Menard et al. 2012 in prep. for full details of the algorithm.). The PRIMUS galaxies induce a correlation on the LBGs through magnification, however, this correlation will nulled out when plotted as an over-density relative to the average density over the whole redshift range as it varies slowly (See Table \ref{tab:mean_values}) and will therefore not have the same signature as the clustering correlation in this test.

\begin{figure}
\begin{center}
\includegraphics[width=0.495\textwidth,clip=true,trim=0 3in 0 0]{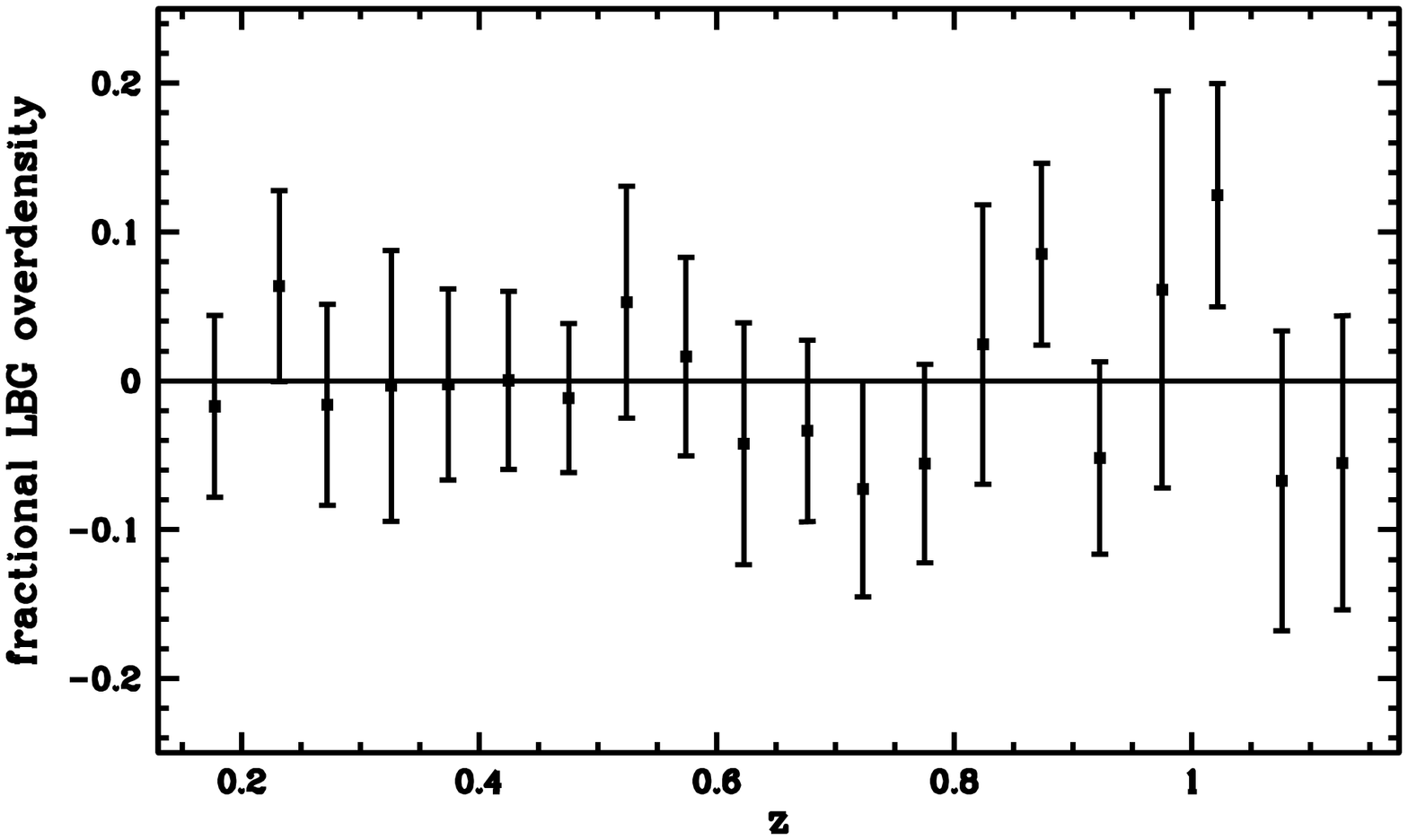}
\end{center}
\caption{\label{fig:redshift_recovery}
Test for low redshift galaxy contamination of the LBG sample, using the PRIMUS spectroscopic redshifts (\citet{Coil11}, Cool et al. in prep.). Plotted is the fractional over-density of LBGs in a 100-1000 kpc radial bin around the $z < 1$ spectroscopic objects. If there were a significant number of low redshift galaxy contaminants of the LBG sample, this correlation would show some structure with redshift. As it does not, we can be confident that the LBG sample is free of low redshift contaminants. The induced correlation by magnification of the LBGs by the PRIMUS galaxies will be mostly nulled out as the over-density is computed with respect to the average density over the redshift range plotted and the amount of magnification over this range varies slowly (See Table \ref{tab:mean_values}).}
\end{figure}

If we are truly selecting Lyman Break Galaxies at redshift $z\sim4$ then we should see no significant cross-correlation between these low redshift spectroscopic samples and the candidate LBGs. Figure~\ref{fig:redshift_recovery} shows the mean over-density of LBGs in an annulus surrounding each spectroscopic object in PRIMUS as a function of redshift (error bars are from a spatial jackknife).  The lack of coherent structure above that expected from an uncorrelated  background sample implies that there is no detectable contamination of the LBG sample by low redshift galaxies.

\newpage

\bibliographystyle{mn2e}
\bibliography{dls_magnification_mnras}

\end{document}